\documentclass{raa}
\usepackage{graphicx,times}             %for PS/EPS graphics inclusion, new
\usepackage{natbib}
\usepackage{amssymb,amsmath}
\bibpunct{(}{)}{;}{a}{}{,}
\usepackage{xcolor}

\usepackage[pagebackref=true]{hyperref}
\hypersetup{colorlinks = true, linkcolor = blue, anchorcolor = red, citecolor = blue, filecolor = red, pagecolor = red, urlcolor = red}

\begin{document}

   \title{A study of holographic dark energy models with configuration entropy
%\footnotetext{$*$ Supported by the National Natural Science Foundation of China.}
}
%   \subtitle{I. Place Your Subtitle Here}

   \volnopage{Vol.0 (20xx) No.0, 000--000}      %%preserved for Editor. DOn't remove!
   \setcounter{page}{1}          %%starting page, preserved for Editor. DOn't remove!

   \author{Biswajit Das
      \inst{1}
   \and Biswajit Pandey
      \inst{2}
   %\and B. J. Smith
     % \inst{3}
   }

   \institute{Department of Physics, Visva-Bharati University, Santiniketan, 
              Birbhum, 731235, India
             {\it bishoophy@gmail.com}\\
        \and
	     {Departmnet of Physics, Visva-Bharati University, Santiniketan, 
	     Birbhum, 731235, India {\it biswap@visva-bharati.ac.in}}\\
        %\and
             %Full institute address for the third author\\
\vs\no
   {\small Received~~20xx month day; accepted~~20xx~~month day}}

\abstract{The holographic dark energy models provide an alternative
        description of the dark energy. These models are motivated by
        the possible application of holographic principle to the dark
        energy problem. In this work, we present a
          theoretical study of the one parameter Li holographic dark
          energy and the two parameter Barrow holographic dark energy
          models using configuration entropy of the matter
          distribution in the Universe. The configuration entropy
        rate exhibits a distinct minimum at a specific scale factor
        that corresponds to the epoch, beyond which the dark energy
        takes a driving role in the accelerated expansion of the
        Universe. We find that the location of the minimum and
        magnitude of the entropy rate at the minimum are sensitive to
        the parameters of the models. We find the best fit relations
        between these quantities and the parameters of each
        model. We propose that these relations can be
          used to constrain the parameters of the holographic dark
          energy models from the future observations such as the
          SKA. Our study suggests that the signature
            of a large quantum gravitational effect on the future
          event horizon can be detected from the measurements of the
          configuration entropy of the matter distribution at multiple
	  redshifts.
}

   \authorrunning{B. Das \& B. Pandey}            %author_head in even pages
   \titlerunning{Holographic dark energy with entropy}  % title_head in odd pages

   \maketitle
%% The author head (on even pages) and the title head (on odd pages) will be
%% automatically extracted from \author{} and \title{}. Whenever the title is too long,
%% you will be asked to supply a shorter one by inserting either \authorrunning{} or
%% \titlerunning{} before \maketitle. Anyway, you can specify your own heads.
%%
%%
%% Note: In the following text body of your manuscript, please note several differences from
%%       other major journals:
%% (1) \subsection{Please Capitalize the First Letter of Each Notional Word in Subsection Title}
%% (2) Please Capitalize the First Letter of Each Notional Word in all tables' captions

%
%________________________________________________ sections below
%
\section{Introduction}           %% first-level sections will be auto-capitalized
\label{sect:intro}

The current accelerated expansion of the Universe remains one of the
major unsolved problems in Cosmology. It has been confirmed by various
independent observations \citep{riess, perlmutter, komatsu, ade} that
the present Universe is going through a phase of accelerating
expansion which started in the recent past. The observed accelerated
expansion is counter-intuitive due to the presence of matter in the
Universe and attractive nature of gravity. It is important to
understand the driving mechanism which governs the cosmic
acceleration. It is conjectured that a hypothetical component termed
as dark energy is responsible for this acceleration. The simplest
possible candidate for dark energy is the cosmological constant,
denoted as $\Lambda$ in Einstein's equations of general
relativity. The resulting model is termed $\Lambda$CDM model. This model
has been successful in explaining a large number of observations and
is currently considered to be the most favoured model of our
Universe. However, this model does not provide any insights into the
physical origin of dark energy. Some recent observations points out to
a tension with the $\Lambda$CDM model. One of them is the famous $H_0$
tension. Some recent works \cite{lusso, rezaei} concluded that the 
$\Lambda$CDM model is not the best fit to some dataset.

Various alternatives to $\Lambda$ such as k-essence \cite{mukhanov},
rolling scalar field \cite{ratra, caldwell} have been proposed that
introduce a modification in the matter sector of Einstein's equations.
Other alternatives such as $f(R)$ gravity \cite{buchdahl} and
scalar-tensor theory \cite{bransdicke} modify the geometric side of
the field equations of general relativity and are known as modified
gravity theories.  A detailed account of various alternative dark
energy models can be found in 
\cite{copeland, amendola, nojiri, odinstov, bamba}. 
Apart from these two alternatives routes, several other interesting 
proposals have been put forward in the literature. The backreaction 
\cite{buchert}, large local void \cite{tomita, hunt}, entropic force 
\cite{easson}, entropy maximization \cite{radicella, pavon}, 
information storage in the space-time \cite{paddy, hamsa} and 
configuration entropy of the
Universe \cite{pandey17, pandey19} are to name a few.

One of the most important theoretical developments in the last three
decades has been the holographic principle. It was first proposed by
Gerard 't Hooft \cite{hooft}. Leonard Susskind provided a string theoretic
interpretation for it \cite{susskind} and Juan Maldacena came up with the
idea of AdS/CFT correspondence \cite{maldacena} which have found many
applications in different area of physics.  The holographic principle
states that all information contained in a volume of space can be
found from the boundary of the volume. A review of holographic 
principle and its connection to cosmology can be found in 
\cite{bousso} and \cite{fischler}, respectively. The efforts to connect the
energy density of dark energy to the entropy of horizon and horizon
length leads to the holographic dark energy models. The 
original proposal to use holographic principle to describe dark energy 
came from \cite{cohen}. Soon, a number of works appeared which 
discussed different aspects 
of such an effort \cite{horava, thomas, hsu, li}. A number of works 
explored the possibility to generalize the model and to use holography 
as a potential candidate for inflation 
\cite{nojiri2, paul, odinstov2, odinstov3}.
The use of Tsallis entropy formula \cite{tsallis} instead of the 
Bekenstein entropy 
formula \cite{bekenstein} leads to Tsallis holographic dark energy models 
\cite{agostino, tavayef, mohammadi}. A number of works explored the 
interacting Ricci holographic dark energy models \cite{chimento, 
chimento2, chimento3, chimento4, chimento5}. Recently proposed 
Barrow entropy formula \cite{barrow} has led to the formulation of Barrow 
holographic dark energy model \cite{saridakis} which has found many 
applications in cosmology \cite{saridakis2, dixit, setare, huang, 
chakraborty, bhardwaj, srivastava, sheykhi, abreu, pradhan, sharma, 
shikha, adhikary, mamon}.

Pandey \cite{pandey17} proposed that growth of structure in the
Universe starting from the initial smooth stage leads to dissipation
of configuration entropy. Since the rate at which structure grows
depends on the cosmological model concerned, the evolution of entropy
might be helpful to discern one cosmological model from another. It
has been proposed that evolution of configuration entropy can be used
to distinguish different equations of state of dynamical dark energy
\cite{das1}, to determine the mass density parameter and cosmological
constant \cite{das2}, to constrain the parameters of the equation of
state of dynamical dark energy \cite{das3}, to determine the
functional form of large scale linear bias of neutral Hydrogen (HI)
distribution \cite{das4}.  In this work, we consider some holographic
dark energy models and study the evolution of configuration entropy in
those models. We study how the evolution of entropy rate depends on
the model and explore if the values of the model parameters can be
constrained from the evolution of entropy.

\begin{figure}
        \includegraphics[width=7.5cm]{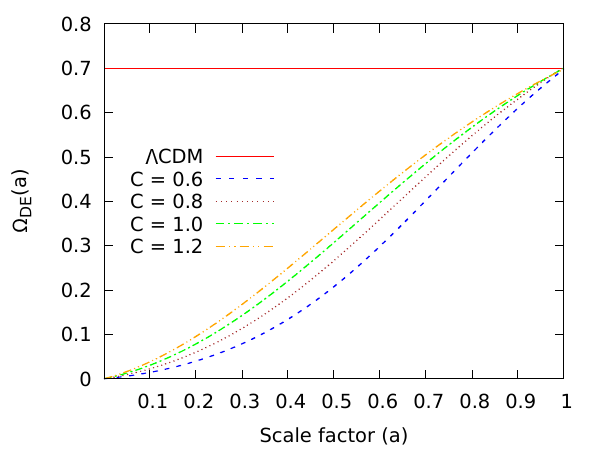}
        \includegraphics[width=7.5cm]{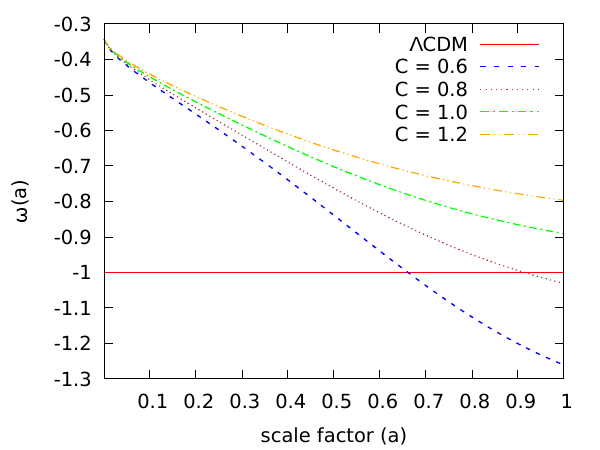}\\
        \includegraphics[width=7.5cm]{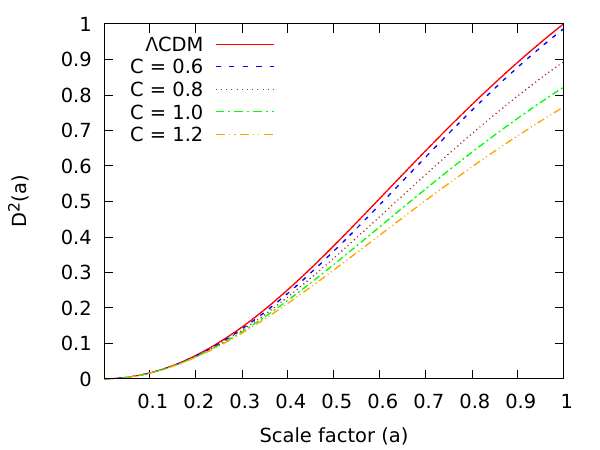}
        \includegraphics[width=7.5cm]{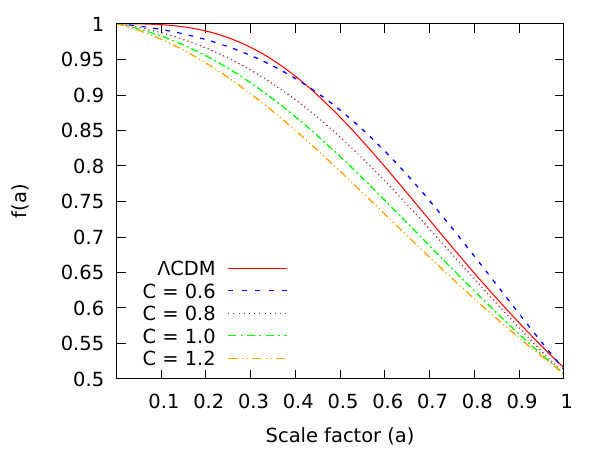}
        \caption{ The top left panel shows the variation of
        $\Omega_{DE} (a)$ with scale factor for different values of $C$ for
        Li holographic dark energy model along with $\Lambda$CDM model.
        The top right panel shows the evolution of $\omega (a)$ with scale
        factor for the same models. The bottom left panel shows the evolution of
        $D^2(a)$ with scale factor for the same models. The evolution of $f(a)$
        with scale factor is shown in the bottom right panel.}
    	\label{fig:one}
\end{figure}

%% Authors can give a citation as 'Michel et al. 1992'.
%% You may also use \cite, \citep and \citet for citation, and use Table~1 or Figure~1
%% and so forth. Using \ref and \label for cross-references of Tables/Figures
%% is a good way in adjusting/adding/removing text, tables or figures.

\section{Theory}
\subsection{Evolution of configuration entropy}

Observations suggest that the Universe is homogeneous and isotropic at 
large scales. But the Universe is highly inhomogeneous and anisotropic
at small scales due to formation of non-linear structures.  We choose
a large enough comoving volume $V$ of the Universe such that the
Universe is nearly homogeneous and isotropic at that length scale. We
divide the volume into subvolumes $dV$. In each of these subvolumes,
we denote the density of matter as $\rho(\vec{x}, t)$. The density is
usually defined at the centre of the subvolume having comoving
coordinate $\vec{x}$ with respect to arbitrary origin and the density
may change with time.  In such a case, we can consider the matter
density field as a random field and we can define the configuration
entropy of the matter density field, following \cite{shannon}, as
~\cite{pandey17}
\begin{eqnarray}
	S_c (t) = - \int \rho(\vec{x}, t) \log \rho(\vec{x}, t) dV.
	\label{eq:one}
\end{eqnarray}

The matter distribution in the Universe can be treated as an ideal
fluid to a good approximation. The continuity equation of that fluid
in an expanding Universe is
\begin{eqnarray}
	\frac{\partial \rho(\vec{x}, t)}{\partial t} + 3 \frac{\dot a}{a} \rho(\vec{x}, t) + \frac{1}{a} \nabla \cdot (\rho(\vec{x}, t) \vec{v}) = 0.
	\label{eq:two}
\end{eqnarray}
Here, $a$ is the cosmological scale factor and $\vec{v}$ is the
peculiar velocity of the fluid element contained in $dV$.  Combining
\autoref{eq:one} and \autoref{eq:two} we get the evolution equation of
configuration entropy as
\begin{eqnarray}
	\frac{dS_c(t)}{dt} + 3 \frac{\dot a}{a} S_c(t) - \frac{1}{a} \int \rho(\vec{x}, t) (3 \dot a + \nabla \cdot \vec{v}) dV = 0.
	\label{eq:three}
\end{eqnarray}
%\textcolor{blue}{
To arrive at \autoref{eq:three}, we note that
\begin{equation}
  \frac{dS_c(t)}{dt} = - \int (1 + \log \rho(\vec{x}, t)) \frac{\partial \rho(\vec{x}, t)}{\partial t} dV.
  \label{eq:twentyone}
\end{equation}
We multiply \autoref{eq:two} by $(1 + \log \rho(\vec{x}, t))$ and integrate over $dV$
to get
\begin{equation}
  \int (1 + \log \rho(\vec{x}, t)) \frac{\partial \rho(\vec{x}, t)}{\partial t} dV + 3
  \frac{\dot a}{a} \int (1 + \log \rho(\vec{x}, t)) \rho(\vec{x}, t) dV + \frac{1}{a} \int (1 + \log \rho(\vec{x}, t)) \nabla \cdot (\rho(\vec{x}, t) \vec{v}) dV = 0.
  \label{eq:twentytwo}
\end{equation}
%\textcolor{blue}{
  The last term in \autoref{eq:twentytwo} is $\frac{1}{a} \int (1 + \log \rho (\vec{x}, t)) \nabla \cdot (\rho(\vec{x}, t) \vec{v}) dV$.
We note that
\begin{equation*}
\nabla \cdot (\rho(\vec{x}, t) \vec{v}) = (\nabla \rho(\vec{x}, t)) \cdot \vec{v} + \rho(\vec{x}, t) \nabla \cdot \vec{v}.
\end{equation*}
So,
\begin{eqnarray*}
  (1 + \log \rho(\vec{x}, t)) \nabla \cdot (\rho(\vec{x}, t) \vec{v}) = \vec{v} \cdot \nabla \rho(\vec{x}, t) (1 + \log \rho(\vec{x}, t)) + \rho(\vec{x}, t) \nabla \cdot \vec{v} (1 + \log \rho(\vec{x}, t))\\
  = \nabla \rho(\vec{x}, t) (1 + \log \rho(\vec{x}, t)) \cdot \vec{v} + \rho(\vec{x}, t) \nabla \cdot \vec{v} + \rho(\vec{x}, t) \log \rho(\vec{x}, t) \nabla \cdot \vec{v}.
\end{eqnarray*}
We write, $\nabla \rho(\vec{x}, t) (1 + \log \rho(\vec{x}, t)) = \nabla (\rho(\vec{x}, t) \log \rho(\vec{x}, t))$.
So,
\begin{eqnarray*}
  (1 + \log \rho(\vec{x}, t)) \nabla \cdot (\rho(\vec{x}, t) \vec{v}) = \nabla (\rho(\vec{x}, t) \log \rho(\vec{x}, t)) \cdot \vec{v} + (\rho(\vec{x}, t) \log \rho(\vec{x}, t)) \nabla \cdot \vec{v} + \rho(\vec{x}, t) \nabla \cdot \vec{v} \\
  = \nabla \cdot (\rho(\vec{x}, t) \log \rho(\vec{x}, t) \vec{v}) + \rho(\vec{x}, t) \nabla \cdot \vec{v}.
  \end{eqnarray*}
The first term on the left hand side of \autoref{eq:twentytwo} is $- \frac{dS_c(t)}{dt}$, the second term gives us $3 \frac{\dot a}{a} \int \rho(\vec{x}, t) dV$ and
$3 \frac{\dot a}{a} \int \rho(\vec{x}, t) \log \rho(\vec{x}, t) dV = - 3 \frac{\dot a}{a} S_c(t)$. We get from the third term $\frac{1}{a} \int \nabla \cdot (\rho(\vec{x}, t) \log \rho(\vec{x}, t) \vec{v}) dV$ and
$\frac{1}{a} \int \rho(\vec{x}, t) \nabla \cdot \vec{v} dV$.
Putting it all together and simplifying, we get
\begin{equation}
  \frac{dS_c(t)}{dt} - 3 \frac{\dot a}{a} \int \rho(\vec{x}, t) dV + 3 \frac{\dot a}{a}
  S_c(t) - \frac{1}{a} \int \rho(\vec{x}, t) \nabla \cdot \vec{v} dV
  - \frac{1}{a} \int \nabla \cdot (\rho(\vec{x}, t)
  \log \rho(\vec{x}, t) \vec{v}) dV = 0.
  \label{eq:twentythree}
\end{equation}
The last term on the left hand side of \autoref{eq:twentythree} can be expressed as a
surface integral. If the volume of integration is chosen to be large, the contribution
from the last term becomes negligible. We cannot express the fourth term in
\autoref{eq:twentythree} as a surface integral because the integrand is a product of
two scalars. So, we are left with \autoref{eq:three}. %}
\begin{figure}
	%\begin{center}
	%\centering
	\includegraphics[width=7.5cm]{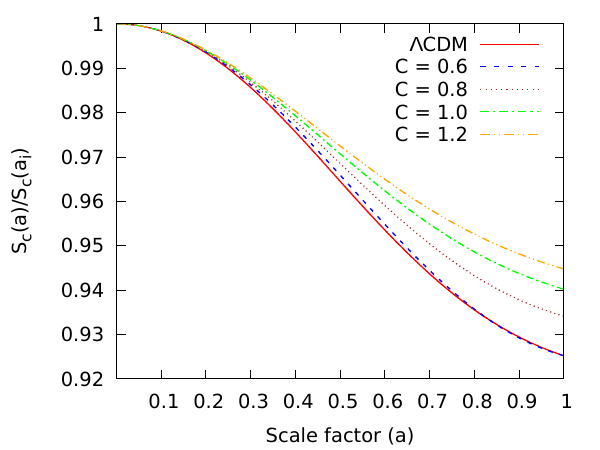}
	\includegraphics[width=7.5cm]{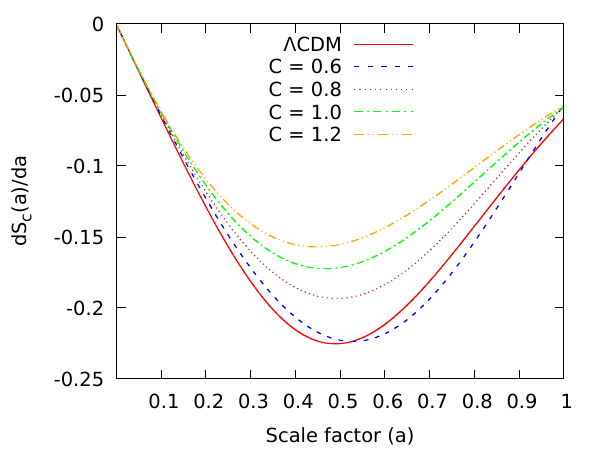}\\
	\includegraphics[width=7.5cm]{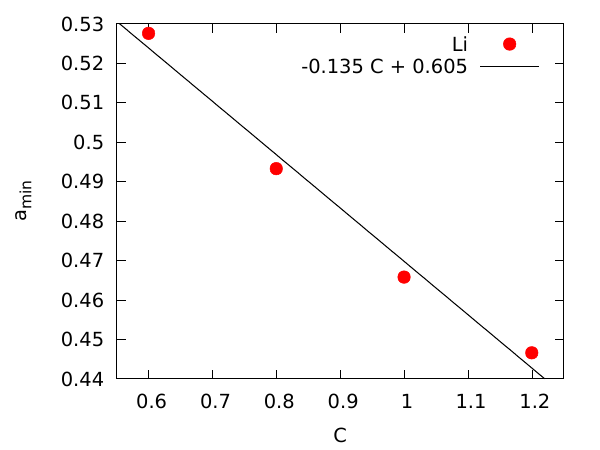}
	\includegraphics[width=7.5cm]{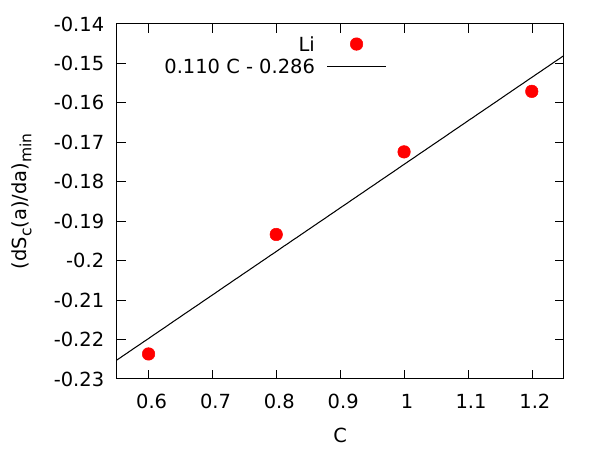}
	\caption{ The top left panel shows the
          evolution of $\frac{S_c(a)}{S_c(a_i)}$ with scale factor for
          different values of $C$ for Li holographic dark energy model
          along with $\Lambda$CDM.  The evolution of
          $\frac{dS_c(a)}{da}$ with scale factor for the same model is
          shown in the top right panel. The bottom left and right
          panels respectively show $a_{min}$ and
          $\left(\frac{dS_c(a)}{da}\right)_{min}$ as a function $C$.
          The best fit straight lines describing these relations are
          shown in the respective panel. }
            \label{fig:two}
	%\end{center}
\end{figure}

\autoref{eq:three} can be rewritten as
\begin{eqnarray}
	\frac{dS_c(a)}{da} \dot a + 3 \frac{\dot a}{a} S_c(a) - 3 M H(a) - \frac{1}{a} \int \rho(\vec{x}, a) \nabla \cdot \vec{v} dV = 0.
	\label{eq:four}
\end{eqnarray}
Here, $H(a)$ is the Hubble parameter, $M$ is the total mass inside the
comoving volume $V$. The variable of differentiation has been changed
from $t$ to $a$. $M = \int \rho(\vec{x}, a) dV = \int \bar \rho (1 +
\delta (\vec{x}, a)) dV$ according to linear perturbation theory where
$\bar \rho = \frac{M}{V}$ is the average density and $\delta (\vec{x},
a) = \frac{\rho(\vec{x}, a) - \bar \rho}{\bar \rho}$ is the density
contrast.  We further simplify Equation \eqref{eq:four} using linear
perturbation theory and get
\begin{eqnarray}
	\frac{dS_c(a)}{da} + \frac{3}{a} \left(S_c(a) - M\right) + \bar \rho \frac{f(a) D^2(a)}{a} \int \delta^2 (\vec{x}) dV = 0.
	\label{eq:five}
\end{eqnarray}
Here, $D(a)$ is the growing mode solution of the evolution equation of
density perturbation in linear approximation and $f(a) = \frac{d \log
  D(a)}{d \log a} = \frac{a}{D(a)} \frac{dD(a)}{da}$ is the
dimensionless linear growth rate. We can integrate \autoref{eq:five} to
get
\begin{eqnarray}
	\frac{S_c(a)}{S_c(a_i)} = \frac{M}{S_c(a_i)} + \left[1 - \frac{M}{S_c(a_i)}\right] \left(\frac{a_i}{a}\right)^3 -
	\left(\frac{\bar \rho \int \delta^2 (\vec{x}) dV}{S_c(a_i) a^3}\right) \int_{a_i}^a da^{\prime} a^{\prime 3} F(a^{\prime}),
	\label{eq:six}
\end{eqnarray}
where $F(a^{\prime}) = \frac{f(a^{\prime}) D^2(a^{\prime})}{a^{\prime}}$, $a_i$ is the initial scale factor and $S_c(a_i)$ is the entropy at $a_i$.
We have chosen $a_i = 10^{-3}$ throughout this work.

To get the evolution of $\frac{S_c(a)}{S_c(a_i)}$ with scale factor we
can either use \autoref{eq:six} or solve \autoref{eq:five}
numerically. One can get the evolution of $\frac{dS_c(a)}{da}$ with
scale factor by simply using \autoref{eq:five}. We require
the knowledge of $D(a)$ and $f(a)$ in a given cosmological model in
order to study the evolution of configuration entropy. We discuss the
evolution of growing mode $D(a)$ and growth rate $f(a)$ in the next
section. During the initial stages of structure formation, $D(a)$ is
very small, so the evolution of $\frac{S_c(a)}{S_c(a_i)}$ in
\autoref{eq:six} is almost entirely determined by $S_c(a_i)$ and
$M$. These quantities do not depend on the cosmological model
concerned and are free parameters of the equation. If $M > S_c(a_i)$,
we expect to see a sudden rise in $\frac{S_c(a)}{S_c(a_i)}$ near $a_i$
whereas $M < S_c(a_i)$ will give rise to a sudden drop. These
variations are due to the choice of initial conditions and we set $M =
S_c(a_i)$ to get rid of them. We also set $\bar \rho \int \delta^2
(\vec{x}) dV = 1$ in \autoref{eq:six} for simplicity.
%\textcolor{blue}{
  An objection regarding the definition of configuration entropy is that while
  Shannon entropy is dimensionless, the configuration entropy has the dimension of
  mass. Hence, the definition of configuration entropy is wrong. However, it is
  easy to make the definition of configuration entropy dimensionless. We can redefine
  the configuration entropy as
  \begin{equation}
    S_c(t) = - \frac{1}{M} \int \rho(\vec{x}, t) \log \rho(\vec{x}, t) dV.
    \label{eq:twentyfour}
  \end{equation}
  Here, $M$ is the total mass inside the comoving volume $V$. If we use
  \autoref{eq:twentyfour} instead of \autoref{eq:one} in our formalism, we obtain
  the differential equation of evolution of configuration entropy as
  \begin{equation}
    \frac{dS_c(a)}{da} + \frac{3}{a} [S_c(a) - 1] + \frac{\bar \rho}{M} \frac{f(a) D^2(a)}{a} \int \delta^2(\vec{x}) dV = 0.
    \label{eq:twentyfive}
  \end{equation}
  We can compare \autoref{eq:twentyfive} and \autoref{eq:five} to find that they differ
  by some constant. Since we are interested in the temporal evolution of $S_c(a)$, we
  may as well set the values of these consants to $1$. In that case, the evolution
  of entropy in \autoref{eq:five} and \autoref{eq:twentyfive} becomes exactly equal.
 % }

%
%               one-column-spanning table
%________________________________________ Table 2: Use_of_the routines

%%Please Capitalize the First Letter of Each Notional Word in table's caption

%%%%%%%%%%%%%%%%%%%%%%%%%%%%%%%%%%%%%%%%%%%%%%%%%%%%%%%%%%%%%%
%%     Examples for figures using graphicx for LaTeX 2e
%%               -- our recommended way for embodying graphics
%%%%%%%%%%%%%%%%%%%%%%%%%%%%%%%%%%%%%%%%%%%%%%%%%%%%%%%%%%%%%%
%
%      A figure as large as the width of the column
%-------------------------------------------------------------
   
\subsection{Growth rate of density perturbations}

Observations of the Cosmic Microwave Background Radiation (CMBR) over
the past few decades have revealed that the CMBR is very homogeneous
and isotropic. But the same observations also revealed the existence
of very small inhomogenieties in the CMBR maps. It is believed that
these inhomogeneities corresponds to the primordial density
perturbations in the matter sector which were amplified by the
mechanism of gravitational instability over time leading to the
present day structures. When $\delta (\vec{x}, t) << 1$, the evolution
of $\delta(\vec{x}, t)$ with time can be described by a differential
equation as
\begin{eqnarray}
	\frac{\partial^2 \delta (\vec{x}, t)}{\partial t^2} + 2 H(a) \frac{\partial \delta(\vec{x}, t)}{\partial t} - 
	\frac{3}{2} \Omega_{m0} H^2_0 \frac{1}{a^3} \delta (\vec{x}, t) = 0.
	\label{eq:seven}
\end{eqnarray}
Here $H_0$ is the present value of Hubble parameter and $\Omega_{m0}$
is the present value of matter density parameter. We change the
variable of differentiation from $t$ to $a$ and introduce the
deceleration parameter $q(a) = - \frac{a \ddot a}{{\dot a}^2}$ to get
\cite{linderjenkins}
\begin{eqnarray}
	\frac{\partial^2 \delta(\vec{x}, a)}{\partial a^2} + \left(\frac{2 - q(a)}{a}\right) \frac{\partial \delta(\vec{x}, a)}{\partial a} - 
	\frac{3}{2} \frac{1}{a^2} \Omega_{m0} \delta (\vec{x}, a) = 0.
	\label{eq:eight}
\end{eqnarray}

\autoref{eq:eight} can be rewritten as \cite{linderjenkins} 
\begin{eqnarray}
	\frac{d^2D(a)}{da^2} + \frac{3}{2a} \left[1 - \frac{\omega(a)}{1 + X(a)}\right]\frac{dD(a)}{da} - 
	\frac{3}{2} \frac{X(a)}{1 + X(a)} \frac{D(a)}{a^2} = 0, 
	\label{eq:nine}
\end{eqnarray}
where we have used the fact that in linear perturbation theory $\delta
(\vec{x}, a) = d(a) \delta (\vec{x})$. $d(a)$ is the growing mode and
$\delta (\vec{x})$ is the initial density perturbation at
$\vec{x}$. $D(a) = \frac{\delta (\vec{x}, a)}{\delta (\vec{x}, a_i)} =
\frac{d(a)}{d(a_i)}$, $\omega(a)$ is the time dependent equation of
state of dark energy and $X(a) = \frac{\Omega_{m0}}{1 - \Omega_{m0}}
e^{- 3 \int_a^1 \omega(a^{\prime}) d \log a^{\prime}}$. We normalise
$D(a)$ such that at present day scale factor $a_0$, $D(a_0) = 1$ for
$\Lambda$CDM model. Solving \autoref{eq:nine} numerically, we can then find
evolution of growth rate with scale factor.

To get $f(a)$ we use 
\begin{eqnarray}
	f(a) = \left[\frac{\Omega_{m0} a^{-3}}{E^2(a)}\right]^{\gamma},
	\label{eq:ten}
\end{eqnarray}

where $E^2(a) = \left(\frac{\Omega_{m0} a^{-3}}{1 - \Omega_{DE} (a)}\right)^{\frac{1}{2}}$ \cite{salzano} 
($\Omega_{DE} (a)$ is the energy density of the dark energy.)
and $\gamma = 0.55 + 0.05 [1 + \omega(a = 0.5)]$ \cite{linder}. 
For simplicity, we have considered the Universe to have only matter 
and dark energy and no interaction between them.

\subsection{Holographic dark energy models}

\subsubsection{Li holographic dark energy}
If we imagine that our Universe has a characteristic length scale $L$ and horizon entropy $S$, then \cite{cohen}
\begin{eqnarray}
	\rho_{de} \propto S L^{-4} = 3C^2M^2_p L^{-2},
	\label{eq:eleven}
\end{eqnarray}
where $M_p = \left(\frac{1}{8 \pi G}\right)^{\frac{1}{2}}$ is the reduced Planck mass and $G$ is Newton's constant and  $S \propto L^2$
according to Bekenstein formula \cite{bekenstein}.

The simplest choice for $L$ is $L = \frac{1}{H(a)}$. In this case the energy density is comparable to present day dark energy energy density 
\cite{horava, thomas} but the equation of state is wrong \cite{hsu}
(\cite{zimdahl} points out that interaction between dark energy and dark matter can give rise to accelerating expansion with Hubble horizon as infra-red cut-off.). 
The particle horizon as $L$ does not produce accelerated expansion \cite{li}. The choice of future event horizon as $L$, 
$L = a \int_t^\infty \frac{dt^{\prime}}{a} = a \int_a^\infty \frac{da^{\prime}}{H(a) a^{\prime 2}}$ 
gives a model of accelerating Universe \cite{li} with correct equation of state.
The density parameter of dark energy in this model satisfies the following equation \cite{wang}, 
\begin{eqnarray}
	\frac{d\Omega_{DE}}{da} = \frac{1}{a} \Omega_{DE} (1 - \Omega_{DE}) \left[1 + \frac{2 \Omega^{\frac{1}{2}}_{DE}}{C}\right].
	\label{eq:twelve}
\end{eqnarray}
\autoref{eq:twelve} can be used to find $\Omega_{DE}$ as function of $a$ and we can use that knowledge to get $E^2(a)$. 
We have chosen the initial condition of \autoref{eq:twelve} such that $\Omega_{DE} (a_0) \sim 0.7$. The equation of state is given by \cite{wang} 
\begin{eqnarray}
	\omega(a) = - \frac{1}{3} - \frac{2 \Omega^{\frac{1}{2}}_{DE}(a)}{3C}.
	\label{eq:thirteen}
\end{eqnarray}
The model has one free parameter, $C$. We can choose different values of 
$C$ to get different evolution of $\Omega_{DE} (a)$ and $\omega(a)$. 
A number of works has constrained the value of the free parameter 
$C$  to be less than $1$ \cite{huang2, zhang, chang, zhang2, ma, xu, xu2, lili, geng}. In this work 
we have chosen four values of $C$ given by $0.6$, $0.8$, $1.0$ and $1.2$.

\begin{figure}
	%\begin{center}
	%\centering
	\includegraphics[width=7.5cm]{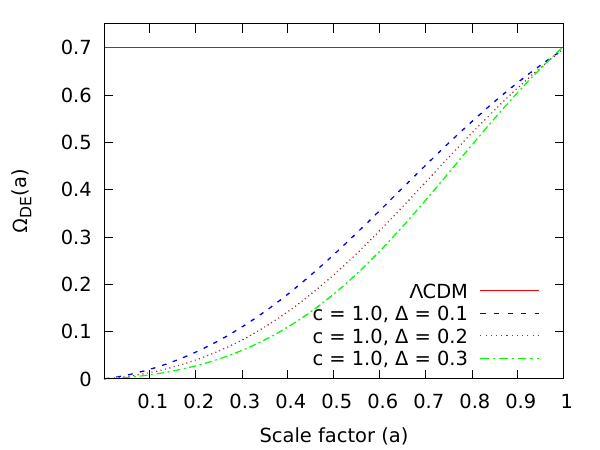}
	\includegraphics[width=7.5cm]{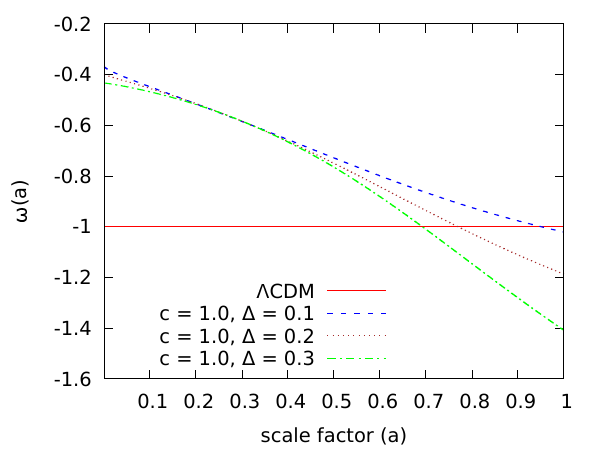}\\
	\includegraphics[width=7.5cm]{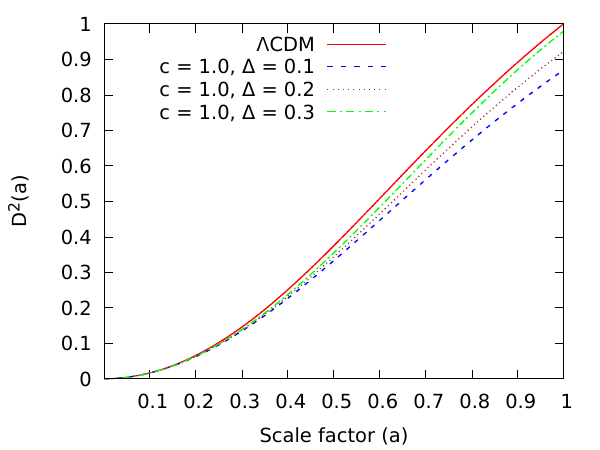}
	\includegraphics[width=7.5cm]{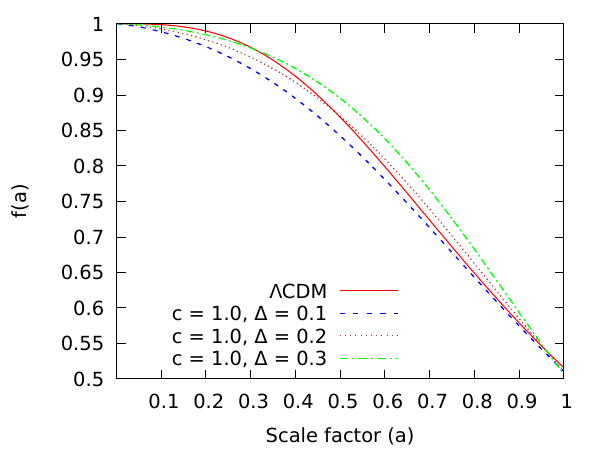}
	\caption{ The same as \autoref{fig:one} but
          for Barrow holographic dark energy model.  In all the models
          we have fixed $c = 1.0$.}
        \label{fig:three}
	%\end{center}
\end{figure}

\begin{figure}
	%\begin{center}
	%\centering
	\includegraphics[width=7.5cm]{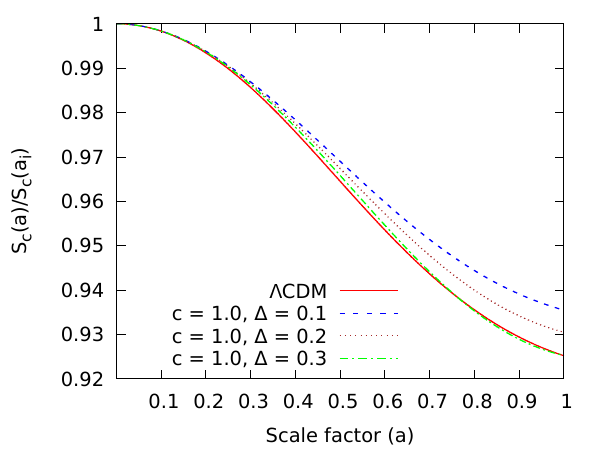}
	\includegraphics[width=7.5cm]{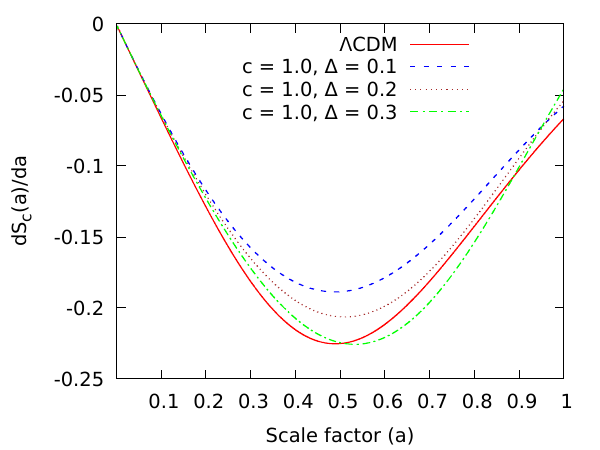}
	\caption{The same as the top two panels of
          \autoref{fig:two} but for Barrow holographic dark energy
          model with a fixed value of $c = 1.0$. }
        \label{fig:four}
	%\end{center}
\end{figure}

\begin{figure}
	%\begin{center}
	%\centering
	\includegraphics[width=7.5cm]{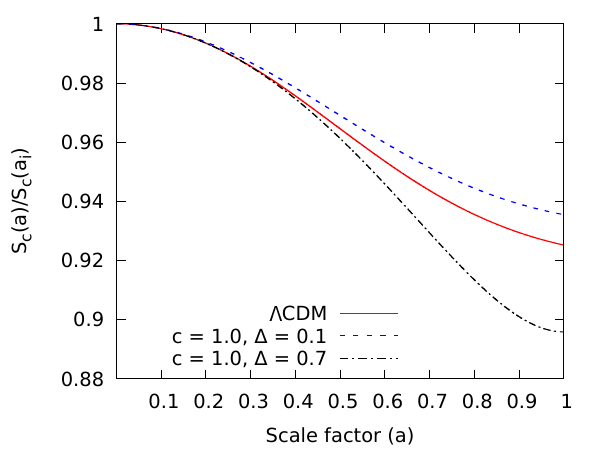}
	\includegraphics[width=7.5cm]{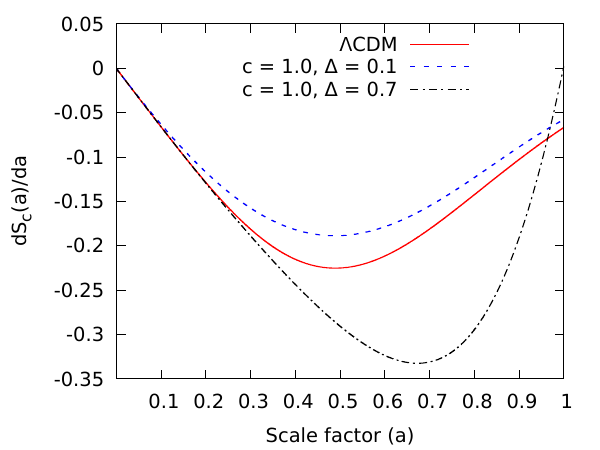}
	\caption{ The same as \autoref{fig:four}
          but for a wider variation of $\Delta$ in Barrow holographic
          dark energy model with a fixed value of $c = 1.0$. }
        \label{fig:five}
	%\end{center}
\end{figure}

\subsubsection{Barrow holographic dark energy}

\begin{figure}
	%\begin{center}
	%\centering
	\includegraphics[width=7.6cm]{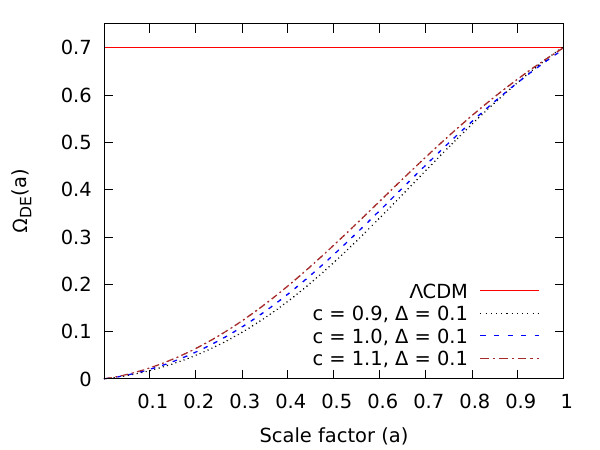}
	\includegraphics[width=7.5cm]{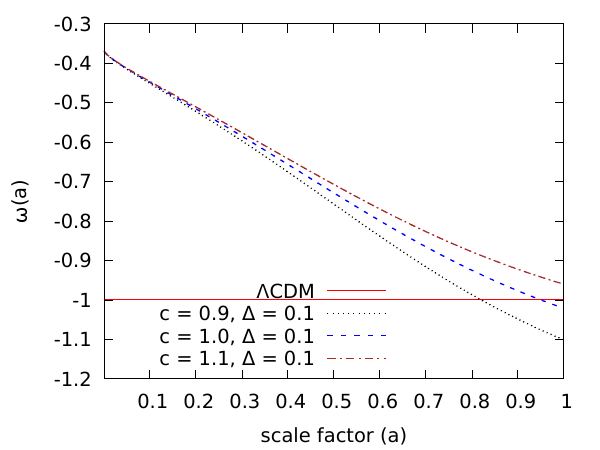}\\
	\includegraphics[width=7.5cm]{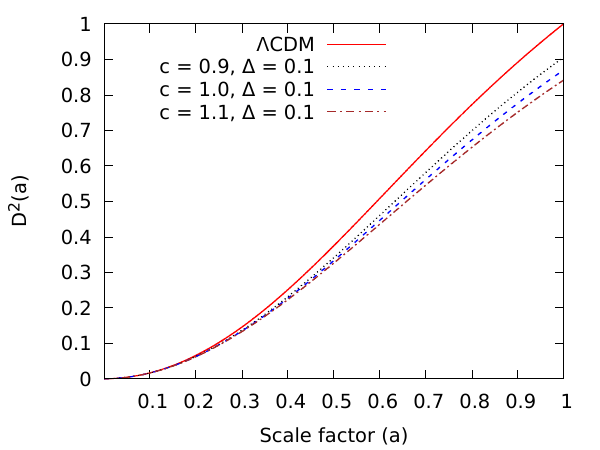}
	\includegraphics[width=7.5cm]{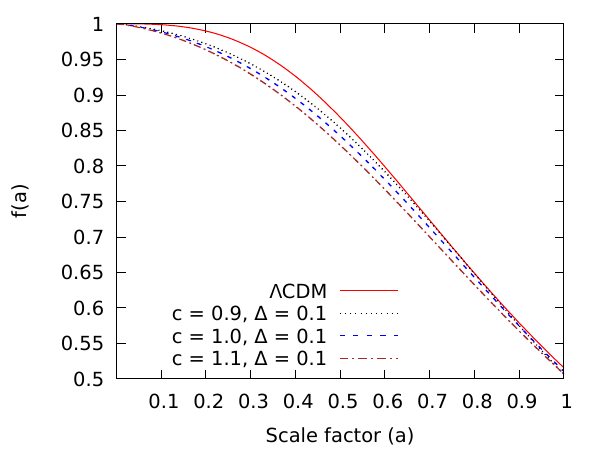}
	\caption{ The same as \autoref{fig:three}
          but for a fixed $\Delta = 0.1$}
        \label{fig:six}
	%\end{center}
\end{figure}

\begin{figure}
	%\begin{center}
	%\centering
	\includegraphics[width=7.5cm]{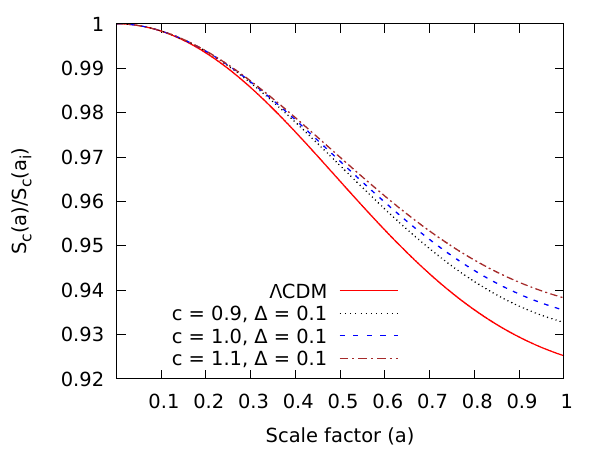}
	\includegraphics[width=7.5cm]{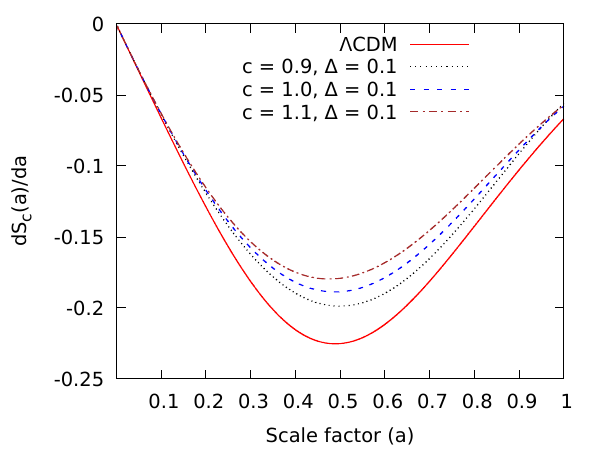}
	%\resizebox{7.5cm}{!}{\includegraphics{secderiv_barrow_del01.eps}}
	\caption{ The same as \autoref{fig:four}
          but for a fixed value of $\Delta = 0.1$.}
        \label{fig:seven}
%\end{center}
\end{figure}

Recently it has been proposed that quantum gravitational effects may
lead to a wrinkled horizon of a black hole instead of a smooth
one. Since the Bekenstein-Hawking formula of black hole entropy is
proportional to the horizon area, it is modified in case of quantum
gravitational effects. The Barrow entropy formula replaces the
Bekenstein-Hawking formula in this case, which is given by
\cite{barrow}
\begin{eqnarray}
	S_B = \left(\frac{A}{A_0}\right)^{1 + \frac{\Delta}{2}}.
	\label{eq:seventeen}
\end{eqnarray}
Here $A$ is the area of the black hole horizon and $A_0$ is the Planck
area. The exponent $\Delta$ encapsulates the departure from
Bekenstein-Hawking formula. $\Delta = 0$ implies no quantum effects,
$\Delta = 1$ implies maximum quantum effects.

In standard holographic dark energy, $\rho_{B} L^4 \leq S_B$ where $L$
is the horizon length and $S_B$ is entropy. Using Barrow entropy formula we get \cite{saridakis,
  saridakis2, barrow2, basilakos, salzano}
\begin{eqnarray}
	\rho_{B} =  3c^2M^2_pL^{2(\frac{\Delta}{2} - 1)}, 
	\label{eq:eighteen}
\end{eqnarray}
where $c$ is one of the free parameters. 
We use the future event horizon as the horizon length. 
(though, \cite{salzano} points out that accelerating expansion can be acieved for this 
model using Hubble horizon as well, without the need to introduce interaction between 
dark matter and dark energy.)
The evolution equation of $\Omega_{DE}$ becomes
\cite{saridakis, saridakis2, basilakos, salzano}
\begin{eqnarray}
	\frac{d\Omega_{DE}}{da} = \frac{1}{a} \Omega_{DE} (1 - \Omega_{DE}) \left[\Delta + 1 + Q (1 - \Omega_{DE})^{\frac{\Delta}{2 (\Delta - 2)}}
	\Omega^{\frac{1}{2 - \Delta}}_{DE} a^{\frac{3 \Delta}{2 (\Delta - 2)}}\right], 
	\label{eq:nineteen}
\end{eqnarray}
where $Q = (2 - \Delta) c^{\frac{2}{(\Delta - 2)}} \left(H_0 \Omega^{\frac{1}{2}}_{m0}\right)^{\frac{\Delta}{2 - \Delta}}$.
The equation of state is given by \cite{saridakis, saridakis2, basilakos, salzano}
\begin{eqnarray}
	\omega(a) = - \frac{1 + \Delta}{3} - \frac{Q}{3} \Omega^{\frac{1}{(2 - \Delta)}}_{DE} (1 - \Omega_{DE})^{\frac{\Delta}{2 (\Delta - 2)}}
	a^{\frac{3 \Delta}{2 (2 - \Delta)}}.
	\label{eq:twenty}
\end{eqnarray}
\autoref{eq:twenty} reduces to Li holographic model for $\Delta = 0$.
We have used $M_p = 1$ for this model.
A few works which have tried to constrain the values of the 
free parameters of this model are \cite{basilakos, barrow2, salzano}.
 We consider the values $0.9$, $1.0$ and $1.1$ for $c$ and $0.1$, $0.2$ and $0.3$ for 
 $\Delta$.

\section{Results and Conclusions}

\begin{figure}
	%\centering
	\includegraphics[width=7.5cm]{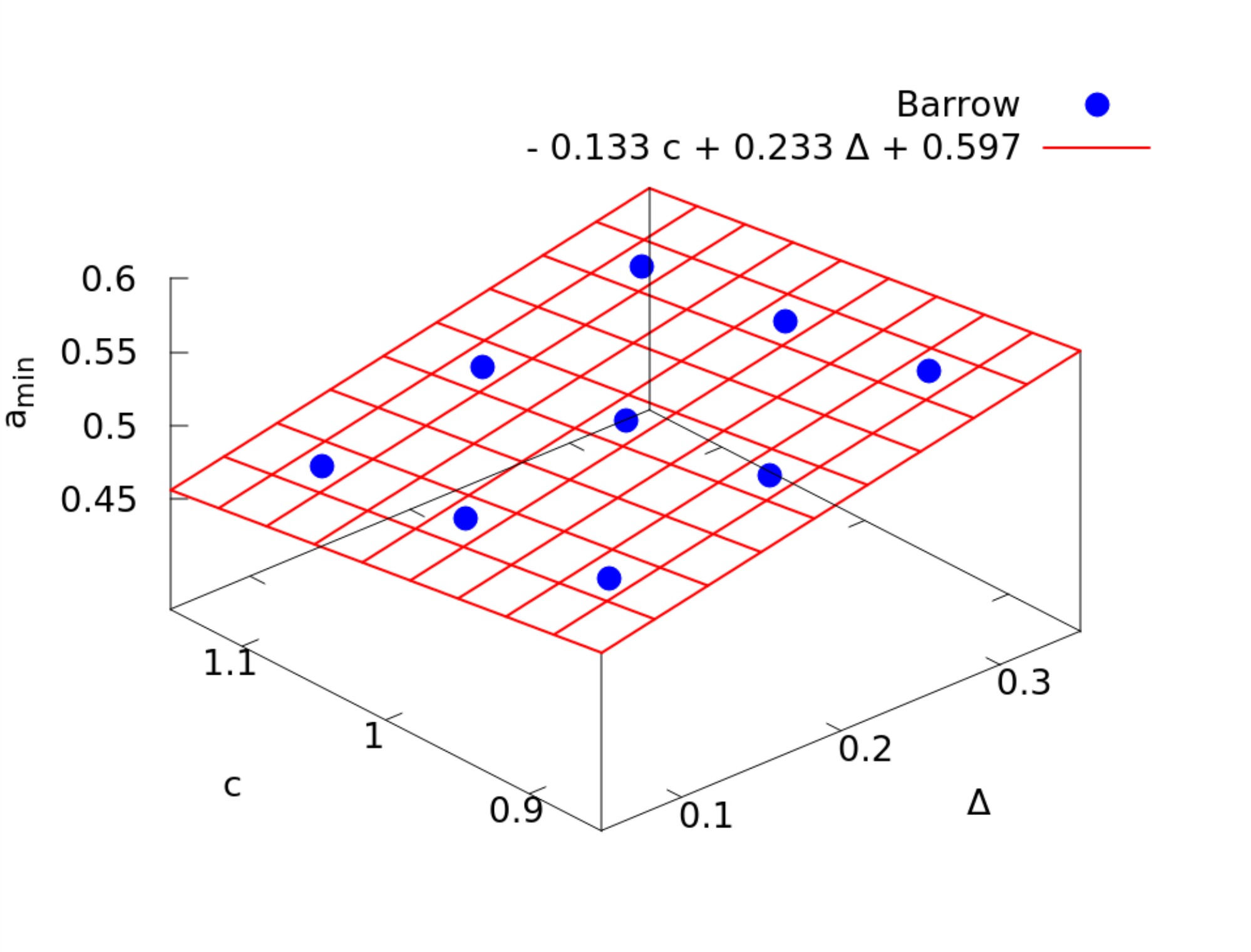}
	\includegraphics[width=7.5cm]{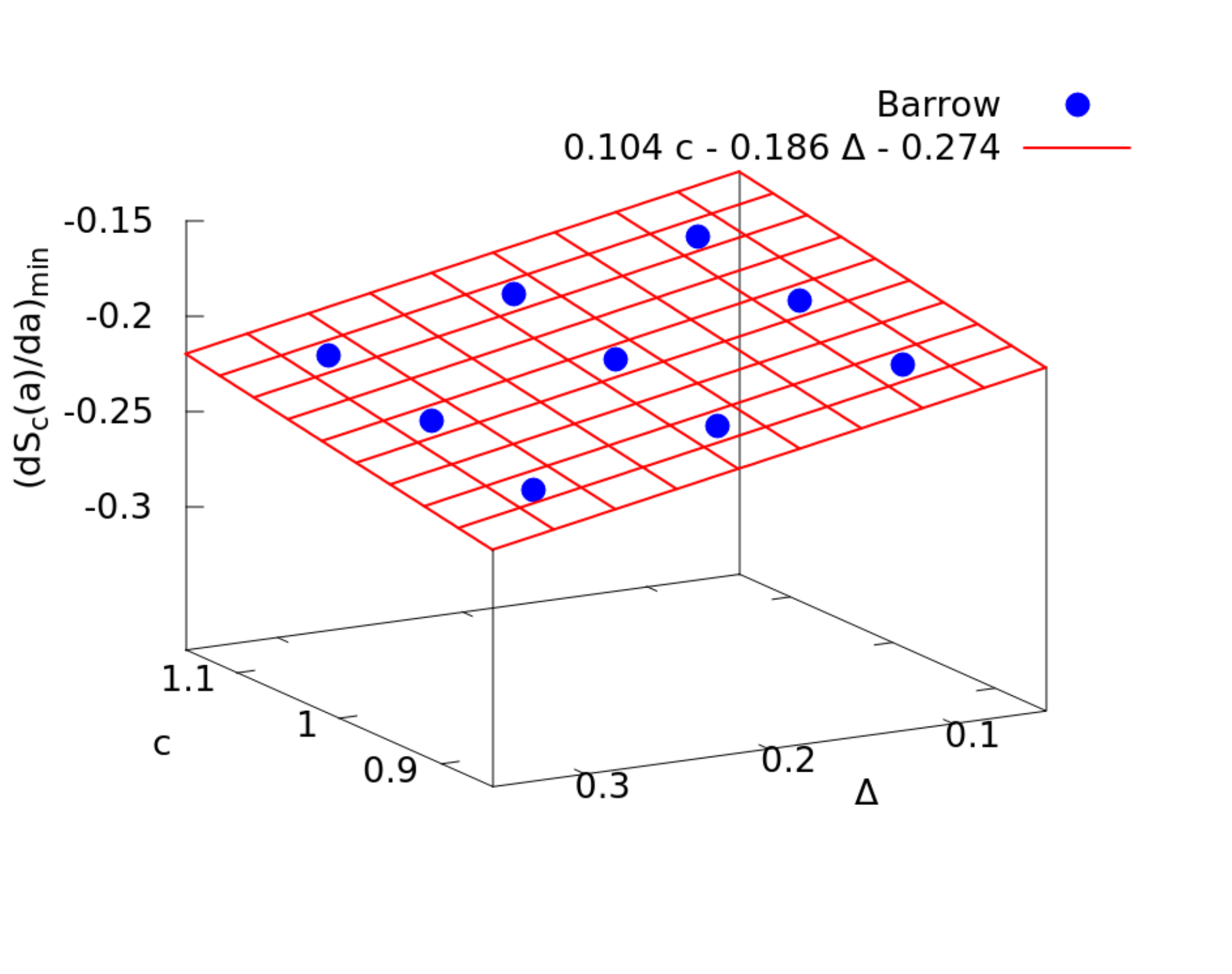}
	\caption{ The left panel shows the dependence
          of $a_{min}$ on the two free parameters $k$ and
	$\Delta$ of Barrow model.  The right panel shows the
          dependence of $\left(\frac{dS_c(a)}{da}\right)_{min}$
          on the same parameters. }
        \label{fig:eight}
\end{figure}
We show the results for the Li holographic dark energy model in
\autoref{fig:one} and \autoref{fig:two}.  In the top left panel
of \autoref{fig:one} we show the variation of $\Omega_{DE} (a)$
with scale factor and the top right panel shows the evolution of
$\omega(a)$ with scale factor. The evolution of $D^2(a)$ and $f(a)$
with scale factor are shown in the bottom left and right panels of
\autoref{fig:one}, respectively. The results for the $\Lambda$CDM
model are also shown together in each panel of \autoref{fig:one}
for comparison.

We show the evolution of $\frac{S_c(a)}{S_c(a_i)}$ and
$\frac{dS_c(a)}{da}$ with scale factor in the top left and right
panels of \autoref{fig:two}, respectively. Since evolution of
entropy is determined by the second and third term in the right hand
side of \autoref{eq:six} and our choice of initial conditions forces the
second term to vanish, the evolution is determined by the third term
which includes a product of $f(a)$ and $D^2(a)$. The top two panels of
\autoref{fig:two} show that initially entropy decreases and the
entropy rate $\frac{dS_c(a)}{da}$ becomes more negative with
increasing scale factor. The decay in the entropy rate continues upto
a a scale factor of $a \sim 0.5$. The rate of decrease of entropy slow
down for all $C$ values after $a \sim 0.5$. The entropy rate then
starts to grow while remaining negative, which signifies a slower
dissipation of the configuration entropy with time. We denote the
scale factor at which the entropy rate turns around as $a_{min}$ and
the magnitude of $\frac{dS_c(a)}{da}$ at $a_{min}$ as
$\left(\frac{dS_c(a)}{da}\right)_{min}$. We calculate $a_{min}$ and
$\left(\frac{dS_c(a)}{da}\right)_{min}$ for different values of the
parameter $C$ in the Li model. We then determine the best fit straight
lines to the numerically obtained values of these quantities in terms
of the parameter $C$. The best fit relations describing these
quantities in the Li model are shown in the bottom two panels of
\autoref{fig:two}.

The Barrow holographic dark energy model is a two
  parameter model and we would like to explore the evolution of
  configuration entropy and entropy rate for different possible
  combinations of the two parameters $c$ and $\Delta$. To better
understand the effect of each parameter on the evolution of each
quantities, we varied one of the parameters while keeping the other
fixed.  This resulted in two sets of plots for Barrow holographic dark
energy models.

We first show $\Omega_{DE} (a)$, $\omega(a)$, $D^2(a)$ and $f(a)$ for
a fixed value of $c$ but for different values of $\Delta$ in this
model in \autoref{fig:three}. Here we fixed $c=1.0$ and allow
$\Delta$ to vary. We find that the equation of state $\omega(a)$
strongly depends on the value of $\Delta$. We also show together the
results for the $\Lambda$CDM model in each panel of
\autoref{fig:three}.

In the left and right panels of \autoref{fig:four}, we respectively
show the evolution of configuration entropy and entropy rate in the
Barrow model for a fixed value of $c$ and different values of
$\Delta$. We find that the configuration entropy decays with scale
factor in each case. The negative entropy rate turns around a specific
scale factor, which is highly sensitive to the parameter $\Delta$. The
value of $a_{min}$ shifts towards higher scale factors with increasing
values of $\Delta$.  This is a result of strong dependence of equation
of state on $\Delta$. The parameter $\Delta$ represents the
modifications in the area of the horizon due to quantum gravitational
effects. The higher sensitivity of $a_{min}$ and
$\left(\frac{dS_c(a)}{da}\right)_{min}$ to the parameter $\Delta$ in
the Barrow model suggest that it may be possible to identify the
signatures of quantum gravitational effects in the behaviour of
configuration entropy and entropy rate. Since $\Delta = 0$ corresponds
to no quantum effects and $\Delta = 1$ corresponds to maximum quantum
effects, we separately compare the effects of a wider variation of
$\Delta$ in \autoref{fig:five}. The left and right panels of
\autoref{fig:five} show that for $\Delta=0.7$, the configuration
entropy dissipates much faster than the $\Lambda$CDM model and can be
easily discerned from it. The results clearly suggest that the
signature of a large quantum gravitational effect can be identified
from the study of the evolution of configuration entropy.

We then repeat the above analysis for a fixed $\Delta = 0.1$ but for
different values of $c$. Different panels of \autoref{fig:six}
show the variation of $\Omega_{DE}(a)$, $\omega(a)$, $D^2(a)$ and
$f(a)$ with scale factor for Barrow holographic dark energy model with
$\Delta = 0.1$. The results show that these quantities are only mildly
sensitive to $c$. The evolution of configuration entropy and entropy
rate in these models are shown respectively in the left and right
panels of \autoref{fig:seven}. The configuration entropy and
entropy rate show a similar characteristics as observed in
\autoref{fig:two} and \autoref{fig:four}. We note that both
$a_{min}$ and $\left(\frac{dS_c(a)}{da}\right)_{min}$ are weakly
sensitive to $c$.

We use the numerical values of $a_{min}$ for different possible
combinations of the parameters $c$ and $\Delta$ in the Barrow model to
find a best fit relation between $a_{min}$ and these
parameters. Similarly, we also find the best fit relation between
$\left(\frac{dS_c(a)}{da}\right)_{min}$ and the parameters of the
Barrow model. We show the best fit planes for $a_{min}$ and
$\left(\frac{dS_c(a)}{da}\right)_{min}$ in the left and right panels
of \autoref{fig:eight}. The equations for the best fit planes in
each case are shown in the respective panels.

In this work we have considered two different holographic dark energy
models. We obtain the evolution of growing mode and dimensionless
linear growth rate by using the knowledge of the evolution equation of
dark energy density parameter and equation of state. We use these
quantities to calculate the evolution of configuration entropy and
entropy rate in these models. For each of the models there are one or
more free parameters. We study the dependence of configuration entropy
and its time derivative on these parameters.  We find from our
analysis that for all models that we have considered, there is a
specific scale factor upto which the entropy rate continue to
decrease.  The negative entropy rate turns around at a specific scale
factor and thereafter the dissipation rate slows down. The scale
factor at which this occurs for a particular model, depends on the
functional form of the equation of state as well as the values of the
parameters. This particular scale factor corresponds to the era where
dark energy density begins to drive the Universe into a phase of
accelerated expansion.  We also find that at this particular scale
factor, the magnitude of entropy rate is different for different
values of the parameters in a particular model. We find that there
exists simple approximate relations between the scale factor of the
minimum and the magnitude of entropy rate and the values of the
parameters. We propose that by measuring configuration entropy at
different scale factors and finding the scale factor at which the
minimum of the entropy rate occurs, one can constrain the values of
the parameters of a particular dark energy model, provided we assume
that it is the correct description of dark energy. One may ask : why
use configuration entropy for this purpose?  We would like to point
out that although entropy is a derived quantity which depends on
$D^2(a)$ and $f(a)$, $D^2(a)$, $f(a)$ and $\frac{S_c(a)}{S_c(a_i)}$
are smooth functions unlike entropy rate which shows a distinct
minimum. It will be easier to identify the position and magnitude of a
minimum rather that finding difference in smooth curves.

We also note that the signature of any quantum gravitational effects
in the holographic dark energy models are reflected in the evolution
of configuration entropy and its time derivative. Possibility of
detecting any such signature using the large scale structure of the
Universe is certainly interesting. Currently no observational data
sets are available to carry out the proposed analysis. In future,
facilities such as SKA would use the redshifted 21 cm signal to map
the density of neutral Hydrogen over a large redshift range. Our
method may then prove to be useful for the study of holographic dark
energy models.

%\acknowledgments
\section{Acknowledgements}

BP would like to acknowledge financial support from the SERB, DST,
Government of India through the project CRG/2019/001110. BP would also
like to acknowledge IUCAA, Pune for providing support through
associateship programme.

\label{lastpage}

\end{document}